**The Ethics of Hacking: Should It Be Taught?**
Nicole Radziwill, Jessica Romano, Diane Shorter, and Morgan Benton

*Poor software quality can adversely affect application security by increasing the potential for a malicious breach of a system. Because computer security and cybersecurity are becoming such relevant topics for practicing software engineers, the need for educational opportunities in this area is steadily increasing. Universities and colleges have recognized this, and have started to offer programs in cybersecurity. At face value, these new programs may not appear controversial, but developing their curriculum requires answering a complex ethical question: Should programs teach hacking to their students? Even though there are different types of hackers, media reports of cybersecurity incidents tend to reserve the "hacker" label for cyber criminals, which overlooks the value in hacking (and, by extension, teaching students to hack). This article examines the full spectrum of hacking behavior, as well as arguments for and against including hacking in education programs, and recommends that hacking skills be considered an essential component of an education and practice in software quality assurance.*

## INTRODUCTION

The definition of "hacker" has evolved since the beginning of computing. Traditionally, the term was used to refer to a radical programmer who aggressively explored creative solutions to problems. However, Falk (2014) and Prasad (2014) explain that hackers are now generally divided into two categories: "white hat" and "black hat" (a distinction that originated at McAfee Corporation, developer of virus protection software). White hats use their talents to subvert criminal activities, while black hats use their abilities for malicious and illegal purposes. They introduce a third category: the "gray hat" hacker who gains unauthorized access to computing resources for the express purpose of helping organizations identify and resolve security issues. Although this type of hacker is not specifically malevolent, they do gain unauthorized access to the computing resources they feign to exploit, and so operate in an ethical gray area.

Despite their intentions of goodwill, gray hat hacking can have unsettling social consequences. In the spring of 2015, CNN reported that cybersecurity consultant Chris Roberts was detailed after allegedly hacking into the control systems of more than 20 United Airlines commercial flights (Perez 2015). Even though he claimed that his actions were intended solely to raise awareness of critical security issues in the aircraft's software, it is not difficult to imagine the negative consequences that might arise if a malevolent "black hat" hacker attempted the same things.

Security has long been a key aspect of software quality assurance. Khan (2010) characterizes the taxonomy of testing as including tests for correctness (based on features and use cases), performance, reliability, and security. He includes "ethical hacking" at the same level as security auditing, risk assessment, and penetration testing. Others (for example, Alexander 2003) have promoted using "misuse cases" as a mechanism to enhance the sensitivities of software designers. The Certified Software Quality Engineer (CSQE) Body of Knowledge includes security in six different elements of its taxonomy, but does not prescribe the approaches that can or should be taken to ensure secure software (ASQ 2008).

The question, then, is not whether attention to (and education about) security is significant, but whether hacking skills *specifically* should be promoted as a means to develop skills in assurance, application design, and quality assurance.

**ETHICAL THEORIES APPLIED TO HACKING**
Whether "ethical hacking" even exists lies at the core of this question. Falk (2014) examined ethical objections to gray hat hacking using three ethical theories: utilitarianism, Kant's maxims and categorical imperative, and Aristotle's virtue ethics. Utilitarianism considers the outcomes of an action or behavior and asks whether it optimizes pleasure while minimizing pain. Because only black hat hacking seeks to maximize the pain inflicted on other parties (typically technological or political), it is the only variant of hacking considered unethical in terms of utilitarianism.

Kant's maxims establish personal rules for self-conduct that are similarly appreciated by others, and his categorical imperative explores the underlying motivations for actions (whether they are motivated by good or by ill). Using this basis for judgment, the black hat hacker is also the only one demonstrating unethical behavior. However, Kant might consider the actions of the black hat to be ethical, if through malicious intent and illegal channels a hacker seeks to weaken or slow the advance of an enemy (or terrorist).

Aristotle's virtue ethics approach asks whether an actor is exercising his or her virtues (which could also loosely be considered skills or talents) to perform a "morally right action." Clearly, white hats and gray hats both believe their intent is morally right, but in the case of the gray hat, the organization whose resources are being exploited may not feel this way. Perhaps even black hats feel that their actions are morally right as well, particularly if they are taking down an enemy.

Despite the inherent subjectivity of these considerations, Falk (2014) argues that the gray hat hacker is just a black hat in a morally ambiguous state. She recommends that "gray hacking is a morally wrong action and as such should be neither condoned by administrators, managers, or other personnel, nor practiced by well-meaning computer professionals." By extension, since these techniques should not be *practiced*, she would also argue that they should not be *taught*.

## OTHER ARGUMENTS AGAINST TEACHING HACKING

Most hackers don't start off with malevolent intent, or the hopes of gaining information to sell for a profit. After interviewing six black hat hackers, Xu, Hu, and Zhang (2013) discovered that the pursuit of hacking often starts off with innocent motives, such as simply wanting to know more about computers, or being able to modify school computers to allow playing games even though it was against school policy. Hacker groups formed as clubs on campus, starting with students who were just curious, to others who were interested in learning how to steal exam files from professors' computers. Though these hackers started without ill intent, "since they were rarely caught and disciplined, they formed the moral value that as long as they do no harm to others, it is not wrong to benefit themselves" (Xu, Hu, and Zhang 2013).

This is a dangerous precedent to set. Teaching hacking in schools provides young people (who may not have developed their own abilities in ethical reasoning) the tools and the knowledge to access secure networks. With this knowledge, they may stumble into black hat situations and face legal consequences. For example, in April 2014, a Western University student was caught after hacking into the Canada Revenue Agency (the Canadian equivalent of the U.S. Internal Revenue Service). Although he was a very bright student, his ethical reasoning was influenced by a previously established moral value, his lawyer argued. At 14, the student had hacked into his school board's computer systems and had not been punished for it, sending him the message that it was "OK to hack" (Bogart 2014; Ha 2014).

Another problem is that some students view hacking as a game, or as an easy way to achieve personal benefits, even though the damage it can cause is tremendous. Turgeman-Goldschmidt (2008), while exploring the sense of meaning hackers get from their pursuits, noted that they are often motivated by a sense of competition with other developers as well as the highly adrenalized sensations that come from experiencing thrills and fun. In the fall of 2014, a Georgia Tech student was caught altering the public calendar of a rival school's football team to say "Get Ass Kicked by Georgia Tech" on a particular game date. It was not reported, however, whether the computer engineering student learned his hacking skills in class or as a hobby (Johnson 2015). An electrical engineering student at Purdue reported using his hacking skills to change his grades on more than one occasion (Weinstein 2014). Another student used his hacking skills to hijack webcams to catch young women in various stages of undress (Humphries 2008).

The impacts of hacking are sensitive and highly context dependent. Hacking is a "*conscious* activity dependent on specific technical skills, operational knowledge of computers, networks, and advanced technological understanding" (Sharma 2007). Given the current age of moral uncertainty and relativism, Sharma uses cognitive and criminological theories to argue that ethics and the ability to differentiate between ethical and unethical behavior should be introduced to educational programs as early as possible. However, he stops short of recommending that hacking be part of the curriculum.

These examples show that younger students may not have the maturity or ethical reasoning to be entrusted with the tools to access and manipulate secure networks. However, conversely, perhaps this provides an even stronger case for engaging young students in the practice of ethical reasoning by stimulating their interest in hacking.

**THE BENEFITS OF INSTRUCTIONAL HACKING**
Some researchers and practitioners argue for the benefits of including hacking in educational programs, pointing out that as the Internet of Things expands and technology is increasingly embedded in people's daily lives, the threat of cyber attacks also increases. As hackers "develop sophisticated methods to penetrate security efforts, the threats of cybersecurity breaches [will] continue to have a large prominence in the minds of management teams across all types of businesses" (Beckett 2015). These proponents pose the question: How is an organization supposed to effectively defend against hackers when their technical staff knows very little about hacking?

At the same time, the job market for security professionals is on the rise and cybersecurity has become a national priority in the United States. According to a Price Waterhouse Coopers study, the cost of cybersecurity for businesses has doubled in the past four years (Beckett 2015). Attention to these issues is now seen as "a tangible way to contribute to the national security effort. For academia, it is also a changing environment when universities can play a role in enhancing the corporate and military abilities to respond quickly to threats" (Kallberg 2013). The workforce requirements are expected to increase significantly over the next decade; it will be natural for universities to seek targeted ways to make cybersecurity education broadly effective (Fourie et al. 2014).

White hat hacking is already being supported, at least informally, at some schools. For example, students from the Saïd Business School at the University of Oxford are helping find ways to identify and treat Ebola patients in Africa. In cooperation with healthcare workers and experimental versions of vaccines, "software developers are now part of the campaign, putting together novel tools which could save lives" (Baraniuk 2014).

Programs focusing on ethical hacking are also starting to emerge. Pike (2013) interviewed 206 industry professionals in cybersecurity to obtain guidance for "ethical hacking" educational programs, and discovered that support for integrating hacking into the curriculum was unanimous. He also identified several critical success factors for making these programs work. These include modeling ethical behaviors, cultivating social interactions and relationships with white hat interest groups and law professionals, engaging in competitions where ethical reasoning must play a role, ensuring that successful individual and team efforts receive recognition, and providing ongoing skills development for hackers. Some educators have proposed specific principles and methods for integrating technical education with the development of ethical reasoning skills (Bratus, Shubina, and Locasto 2010).

Even though cybersecurity professionals are currently in high demand, specialized programs are limited, and nearly all computer science programs do not require cybersecurity coursework to graduate. Furthermore, effective education will be strongly interdisciplinary, requiring "cyber defense research teams to address not only computer science, electrical engineering, software and hardware security, but also political theory, institutional theory, behavioral science, deterrence theory, ethics, international law, international relations, and additional social sciences" (Kallberg and Thuraisingham 2012). Addressing the emerging national and international challenges will require a broader approach to enhancing exposure, which may not be achieved through dedicated cybersecurity programs.

Regardless of whether or not students are *taught* hacking skills, hackers will still exist. If the practical skills associated with hacking or cybersecurity are not taught in schools, it means that all hackers will be self-taught in their technical skills and will be less likely to devote the same interest and attention to ethics and ethical reasoning in the context of cybersecurity. If hacking becomes a part of educational curricula, at least the schools and universities that include the material will have the ability to teach ethical hacking, promoting conscious ethical practices and lowering the likelihood that students will use their knowledge maliciously.

**CONCLUSION**
The expanding workforce pressure for computer and cybersecurity professionals, the steady increase in overall risk and increase in the potential for risk exposure, and the ethical and practical considerations discussed in this article all suggest a general consensus that ethical hacking should be taught in educational programs. The lingering debate is, if hacking is taught in schools, will it breed cybersecurity professionals who will keep networks and economies safe? Or, will it just give more young people the means to become black hat hackers?

Based on the ethical theories considered earlier, one practical approach to mitigate this risk would be for educational programs to focus on navigating the moral and ethical concerns of gray hat hacking scenarios. "Ethical hackers" would still be educated to have the same deep technical skills and abilities as black hat hackers, and may even perform identical operations. However, they would learn their skills in the context of different hacking scenarios, and be required to explore the complex context of their technical decisions with respect to all stakeholders (including society in general).

Software quality professionals know that security considerations must be factored into the design and development of software applications. Their skills in security could be strengthened even further by developing additional proficiency in hacking techniques while exploring ethical reasoning to fully understand the boundaries between white hat and gray hat situations. There is a need for case study development to support this path.